\documentclass[structabstract]{aa} 
\usepackage{graphicx}
\usepackage{txfonts}
\usepackage{lscape}
\newcommand{\vm}{{\,\times\,}}
\begin{document}

\title{The direction of outflows from filaments: \\constraints on core formation}
\author{S. Anathpindika \and A. P. Whitworth}
\institute{School of Physics \& Astronomy, Cardiff University, 5 The Parade, Cardiff CF24 3AA, Wales, UK \\
\email{Sumedh.Anathpindika@astro.cf.ac.uk} \\ \email{Anthony.Whitworth@astro.cf.ac.uk}}

\date{Received ; accepted }

\abstract
{It is generally presumed that the outflows from a YSO are directed close to its rotation axis (i.e. along its angular momentum vector and orthogonal to any attendant accretion disc). Many YSOs are formed from dense prestellar cores embedded in filaments, and therefore the relative orientations of outflows and filaments may place a useful constraint on the dynamics of core formation.}
{{\bf We explore this possibility, from the viewpoint of what it may tell us about the angular momentum delivered to a core forming in a filament. We stress that we are not here addressing the issue of the relationship of filaments and outflows to the prevailing magnetic field direction, although this is evidently also an interesting issue.}}
{We use data from the literature and the SCUBA archive to estimate the projected angles between 45 observed outflows and the filaments which appear to contain their driving sources. The distribution of these angles is then compared with model predictions, so as to obtain a statistical constraint on the distribution of intrinsic angles between outflows and filaments.}
{Using the limited data available, and neglecting any possible selection effects or correlations between nearby outflows, we infer that the observed outflows have a tendency to be orthogonal to the filaments that contain their driving sources. Specfically, in the cases where the directions of the filaments and outflows are well defined, we infer statistically that $72\,\%$ of outflows are within $45^{\rm o}$ of being orthogonal to the filament, and only $28\,\%$ are within $45^{\rm o}$ of being parallel to the filament.}
{This suggests that the prestellar cores which spawned the YSOs driving the observed  outflows had angular momenta which were approximately orthogonal to the filaments out of which the cores formed. We briefly discuss the implications of this for two proposed core formation mechanisms.}


\maketitle

\section{Introduction}

The substructure within a star-forming molecular cloud often appears to be filamentary, and the dense prestellar cores out of which protostars condense are often embedded within filaments (e.g. Schneider \& Elmegreen 1979; Hatchell et al. 2005; Johnstone \& Bally 2006; Kirk, Johnstone \& Tafalla 2007; Nutter \& Ward-Thompson 2007; Kirk, Ward-Thompson \& Nutter 2007; Muench et al. 2007; Goldsmith et al. 2008; Narayanan et al. 2008). The implication is that the growth of prestellar cores is fed mainly by material flowing in along filaments{\bf, and -- although Hatchell et al. (2005) caution against it -- this interpretation is supported by many numerical simulations, (e.g. Passot, V\'azquez-Semadeni \& Pouquet 1995; Nagai, Inutsuka \& Miyama 1998; Padoan \& Nordlund 1999; Balsara, Ward-Thompson \& Crutcher 2001; Klein, Fisher \& McKee 2001; Gammie et al. 2003; Li, P. et al. 2004; Li, Z. \& Nakamura 2004; Nakamura \& Li 2005; Oishi \& Mac Low 2006; Ciolek \& Basu 2006; Kudoh et al. 2007; Kudoh \& Basu 2008; Hennebelle et al. 2008; Offner, Klein \& McKee 2008).} The highly anisotropic inflow {\bf from a filament onto a core} may then have consequences for the dynamics of core collapse and fragmentation. In particular, the net angular momentum of the inflowing material will strongly influence the orientation of binary orbits and circumstellar accretion discs in the small-$N$ subcluster of protostars forming in a prestellar core.

During the embedded Class 0 and Class I phases of protostellar evolution (e.g. Di Francesco et al. 2007; Ward-Thompson et al. 2007), the orientations of circumstellar accretion discs can be inferred from the directions of the outflows which they drive. An outflow is presumed to be driven by torsional MHD waves propagating along magnetic field lines anchored in a circumstellar disc (e.g. Pudritz \& Norman 1986). Therefore the outflow should be approximately parallel to the rotation axis of the circumstellar disc, i.e. along the angular momentum vector of the material forming the parental core.

It follows that the orientation of an outflow, relative to the filament in which its driving source has been born, may contain important information on the dynamics of core formation. Specifically it constrains the relationship between the inflow forming the core (if we assume that this inflow is concentrated along the filament) and the angular momentum which this material brings with it (if we assume that this angular momentum is oriented along the direction of the outflow). This information can be used to discriminate between different mechanisms for core formation (e.g. Whitworth et al. 1995; Banerjee, Pudritz \& Anderson 2006; Banerjee \& Pudritz 2007).

The present paper is organised as follows. In Section 2, we present the distribution of projected angles, $\gamma$, between observed filaments and the outflows from YSOs that appear to be embedded in the filaments. We show that the distribution of projected angles is compatible with the distribution expected if outflows are approximately orthogonal to filaments; relative to the orthogonal direction, the intrinsic directions of outflows have a standard deviation $\sigma_{\psi}\sim 45^{\rm o}$, which implies that $72\,\%$ of outflows are within $45^{\rm o}$ of being orthogonal to the filament which contains their driving source. In Section 3, we discuss two mechanisms for core formation, (i) gravitational fragmentation of a shock-compressed layer (e.g. Whitworth et al. 1995), and (ii) gravo-turbulent fragmentation (e.g. {\bf Klein, Fisher \& McKee 2001;} Banerjee, Pudritz \& Anderson 2006; Banerjee \& Pudritz 2007{\bf ; Offner, Klein \& McKee 2008}). We {\bf infer} that the first mechanism is more likely to produce the observed distribution{\bf , but this is not a very strong inference, and more observational data are required to make it more robust}. We also discuss the statistical significance of the result, and the possible influence of selection effects. The data used are presented in Appendix A. The distribution of projected angles expected is derived in Appendix B.

\section{The distribution of observed and intrinsic angles between filaments and outflows}

\begin{figure}
\centering
\includegraphics[angle=-90,width=8.5cm]{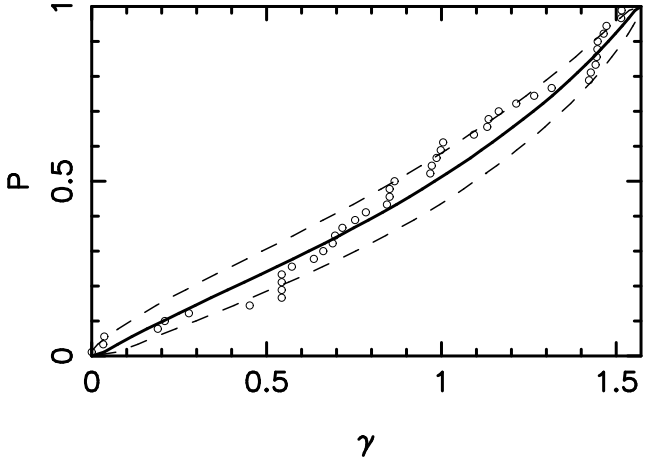}
\caption{The open circles give the cummulative distribution of the observed (i.e. projected) angles $\gamma$ between filaments and outflows. The full line gives the mean cummulative distribution for the best-fitting model. This model has $\sigma_{\psi}=0.8$, where $\psi$ is the intrinsic angle between the outflow and the normal to the filament. The dashed lines give the $\pm 1\sigma$ dispersion about this line.}
\label{FIG:BESTFIT}
\end{figure}

\begin{figure}
\centering
\includegraphics[angle=-90,width=8.5cm]{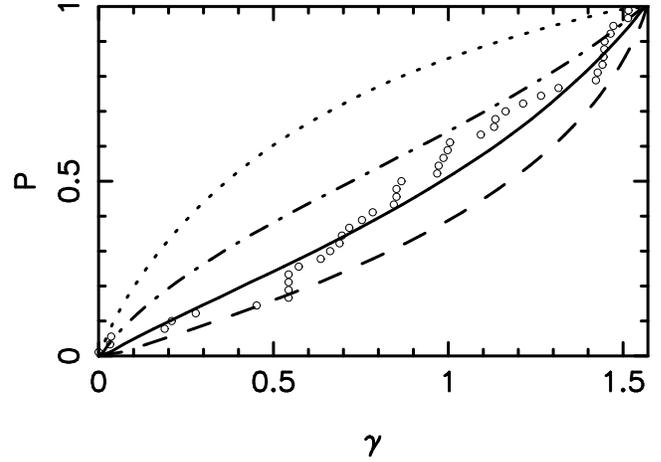}
\caption{The open circles give the cummulative distribution of the observed (i.e. projected) angles $\gamma$ between filaments and outflows. The full line gives the mean cummulative distribution for the best-fitting model ($\sigma_{\psi}=0.8$). The dashed line gives the distribution when all outflows are exactly orthogonal to the filament (i.e. $\sigma_{\psi}=0$). The dot-dash line gives the distribution when all values of $\psi$ are equally likely (i.e. $\sigma_{\psi}=\infty$). The dotted line gives the distribution when the angle between the outflow and the filament (i.e. $\psi' \equiv \pi/2 - \psi$) has a Gaussian distribution with $\sigma_{\psi'}=0.8$.}
\label{FIG:OTHERFIT}
\end{figure}

\begin{figure}
\centering
\includegraphics[angle=-90,width=8.5cm]{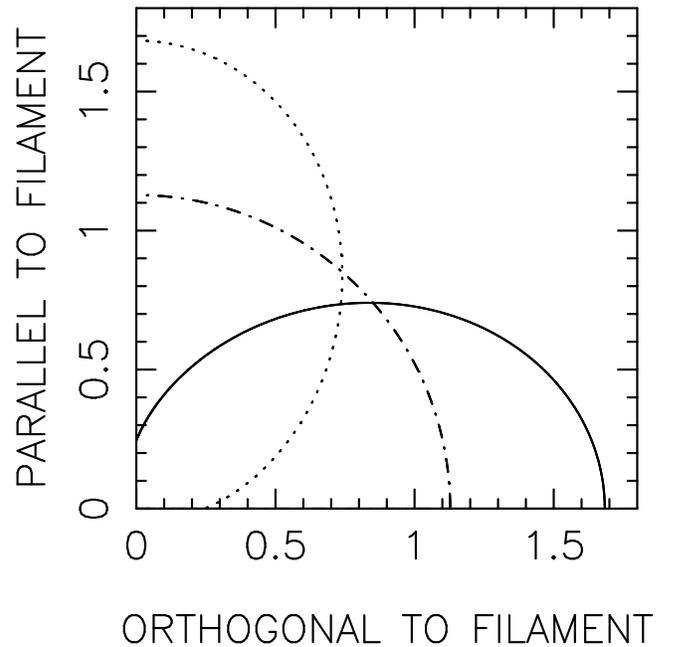}
\caption{Polar diagrams illustrating the distribution of outflow directions (in one quadrant) for the three non-singular cases illustrated on Fig. \ref{FIG:OTHERFIT}. The line types are the same, i.e. full for $\sigma_{\psi}=0.8$, dot-dash for $\sigma_{\psi}=\infty$, and dotted for $\sigma_{\psi'}=0.8$ (where $\psi' \equiv \pi/2 - \psi$). For $\sigma_{\psi}=0.8$, $72\%$ of outflows are within $45^{\rm o}$ of the orthogonal to the filament.}
\label{FIG:POLAR}
\end{figure}

On Figures \ref{FIG:BESTFIT} and \ref{FIG:OTHERFIT}, open circles show the cummulative distribution of $\gamma$, the observed (i.e. {\it projected}) angle between outflows from YSOs that appear to be embedded in filaments and the filaments themselves. The sources of the raw data, and the techniques used to derive the angles, are described in detail in Appendix \ref{APP:OBS}.

To interpret this data, we assume (i) that the {\it intrinsic} angle, $\psi$, between the outflow and the normal to the filament has a truncated Gaussian distribution with standard deviation $\sigma_{\psi}$ (truncated in the sense $0\leq |\psi |\leq\pi/2$); and (ii) that the observed systems are randomly oriented relative to the observer's line of sight. Using this model we have performed, for different values of $\sigma_{\psi}$, a Monte Carlo experiment to generate 20,000 independent random samples, with each sample comprising ${\cal N}=45$ artificial observed systems. Using these random samples, we determine the mean projected angle of the $n^{\rm th}$ member of a sequence, $\mu_{\gamma}(n)$, and its standard deviation, $\sigma_{\!\gamma}(n)$.

On Fig. \ref{FIG:BESTFIT}, the full line shows the cummulative distribution for $\mu_{\gamma}(n)$, and the dotted lines show the cummulative distributions for $\mu_{\gamma}(n)\pm\sigma_{\!\gamma}(n)$, for the best-fit model, with $\sigma_{\psi}=0.8\,{\rm radians}\simeq 45^{\rm o}$. We have identified the best fit by minimizing
\begin{eqnarray}
{\cal Q}&=&\frac{1}{\cal N}\,\sum_{n=1}^{n={\cal N}}\left\{\left(\frac{\gamma(n)-\mu_{\gamma}(n)}{\sigma_{\!\gamma}(n)}\right)^2\right\}\,,
\end{eqnarray}
where $\gamma(n)$ is the $n^{\rm th}$ observed angle; for the best fit, ${\cal Q} \simeq 0.28\,$. $\sigma_{\psi}=0.8$ implies that there is a $72\%$ chance that the outflow is within $45^{\rm o}$ of being {\it orthogonal} to the filament, and hence a $28\%$ chance that the outflow is within $45^{\rm o}$ of being {\it parallel} to the filament.

On Fig. \ref{FIG:OTHERFIT} we show the model distributions ($\mu_{\gamma}(n)$ only) for $\sigma_{\!\psi}=0$ (all outflows exactly orthogonal to the filament; dashed line), $\sigma_{\!\psi}=0.8$ (best fit; full line), and $\sigma_{\!\psi}=\infty$ (no prefered orientation; dash-dot line). In addition we show the results when the angle between the outflow and the filament, i.e. $\psi'\equiv\pi/2-\psi$, has a Gaussian distribution with $\sigma_{\!\psi'}=0.8$ (dotted line). The very large disparity between this last model distribution and the observed one reinforces our conclusion that outflows are very unlikely to be even approximately aligned with filaments.

Fig. \ref{FIG:POLAR} is a polar diagram illustrating the distribution of outflow directions for the three non-singular cases illustrated on Fig. \ref{FIG:OTHERFIT}. The same line types are used, i.e. full for $\sigma_{\psi}=0.8$, dot-dash for $\sigma_{\psi}=\infty$, and dotted for $\sigma_{\psi'}=0.8$ (where $\psi' \equiv \pi/2 - \psi$).

\section{Discussion}

\subsection{Selection Effects}

Our conclusion that outflows tend to be approximately orthogonal to the filaments spawning their driving sources may be corrupted by by two factors.

First, the observed systems (filament, YSO and outflow) are not just small in number, but are also located in a small number of local star formation regions. The orientation of systems in the same star formation region may well be correlated, for example, due to a preferential direction of external compression, or due to a prevailing, approximately uniform, large-scale background magnetic field. Our assumption that the observer's line of sight is randomly oriented relative to each system is then not valid. This problem can probably only be alleviated by extending the observational database. {\bf Indeed, Strom et al. (1986) and Duch{\^e}ne \& M{\'e}nard (2004) find that in the Taurus-Auriga star formation region the outflows -- but not necessarily all the protostellar discs -- tend to be parallel to the large-scale magnetic field.}

Second, it is possible that outflows that propagate approximately orthogonal to a filament are more -- or less -- readily identified than those that propagate approximately parallel to a filament. It is hard to argue rigorously one way or the other. Does the higher density encountered by an outflow propagating approximately parallel to a filament inhibit its advancement and thereby make it harder to identify, or does it increase its visibility by supplying the dense gas required for a dissipative working surface (i.e. an HH Object)? We are inclined to believe the latter, in which case our conclusion should be reinforced, i.e. more than $84\,\%$ of outflows should be within $45^{\rm o}$ of the orthogonal to the filament.

\subsection{Implications for core formation}

In order to speculate on how the apparent preferential orientation of outflows approximately orthogonal to filaments might be interpreted, we consider two possible core formation mechanisms. These two mechanisms are not chosen because they are the only ones that have been proposed, but because, as far as we know, they are the only ones which make specific predictions concerning the relative orientation of filaments and discs (hence outflows).

In the first mechanism (Whitworth et al. 1995), two interstellar gas flows collide and produce a shock compressed layer. In the centre of mass frame of the layer, the collision is unlikely to be head-on, so the collisions partners have net orbital angular momentum, ${\bf H}$, which is orthogonal to the collision velocity, ${\bf v}$. Hence the layer tumbles about $\hat{\bf h}\equiv{\bf H}/|{\bf H}|$. If the post-shock gas is cooler than the pre-shock gas, the layer eventually becomes gravitationally unstable, and breaks up, first into a network of filamets, and then into prestellar cores along the filaments. Because the layer is tumbling about $\hat{\bf h}$, the filaments are also tumbling about $\hat{\bf h}$. Consequently the streamlines delivering material to the core from the two directions of the filament become increasingly offset relative to one another (see Figs. 8 and 9 in Whitworth et al. (1995)) and the prestellar core acquires angular momentum approximately parallel to $\hat{\bf h}$ and orthogonal to the filament. In this case, we might expect the outflows from YSOs formed in the core to be approximately orthogonal to the filament (apparently as observed).

In the second mechanism (e.g. {\bf Klein, Fisher \& McKee 2001;} Banerjee, Pudritz \& Anderson 2006; Banerjee \& Pudritz 2007{\bf ; Offner, Klein \& McKee 2008}), filaments are created in the shear flow between two colliding turbulent streams. In the purely hydrodynamic cases presented by Klein et al. (2001) and Banerjee et al. (2006), and in the magneto-hydrodynamic case presented by Banerjee \& Pudritz (2007), the filament is aligned along the orbital angular momentum of the collision, i.e. along $\hat{\bf h}$. Consequently the filament tends to rotate about its long axis (rather than tumbling). The material flowing into a core from the two directions of the filament now delivers angular momentum which is aligned approximately parallel to the filament. Therefore in this case we might expect the outflows from YSOs formed in the core to be approximately parallel to the filament (in apparent conflict with the observations).

\section{Conclusions}

We have analysed the limited number of protostellar systems in local star formation regions where a YSO embedded in a discernable filament drives an outflow. In these cases, the distribution of angles between filaments and outflows appears to imply that outflows are usually approximately orthogonal to the filaments which contain (and presumably have fed) the cores which have spawned their driving YSOs. Given the chaotic and impulsive nature of multiple star formation in a prestellar core, one should expect some variance in the relative orientation of filaments and outflows. Specifically, it appears that $72\,\%$ of outflows are within $45^{\rm o}$ of orthogonal to the filament which contains their driving YSO. This inference is not very robust, statistically, but, if confirmed, it would favour a model in which filaments were formed by gravitational fragmentation of shock compressed layers, as against one in which filaments were formed from shear instabilities in colliding hydrodynamic or magneto-hydrodynamic flows. Further observations of outflows from filaments are needed to validate this conclusion.

\begin{acknowledgements}
We thank James Di Francesco for his extremely generous cooperation in providing processed data from the SCUBA archives for $\rho$ Ophiuchus, Serpens and NGC1333; David Nutter and Jason Kirk for providing the fully processed data for the OIF and Taurus; Peter Coles and Mike Disney for useful discussions on non-parametric statistical tests; Derek Ward-Thompson and the anonymous referee for useful suggestions. SA gratefully acknowledges postgraduate financial support from The State Government of Maharashtra, India (grant: DSW/Edu/Inf/2004/6283(35)), and from The George Education Trust, London. APW gratefully acknowledges support from The Science and Technology Facilities Council (grant: PP/E000967/1).
\end{acknowledgements}


\appendix

\section{The sample of observed filaments and outflows}\label{APP:OBS}

\begin{table*}
\caption{Observational data (see text for details).}\label{TAB:DATA}
\begin{center}
\begin{tabular}{lllllllrrl} 
\hline
HH Obj & & \hspace{2.5cm} & YSO & & & & \hspace{0.60cm}Angles & & \\
Name & $\alpha_{_{2000}}$ & $\delta_{_{2000}}$ & Name & $\alpha_{_{2000}}$ & $\delta_{_{2000}}$ & Ref. & $\eta_{_{\rm OUT}}\;\;$ & $\eta_{_{\rm FIL}}\;\;\,$ & $\gamma$ \\\hline
&&&&&&&&&\\
\multicolumn{6}{l}{{\bf Orion Integral Filament}$\;\;$(SCUBA 850$\mu$m [1])} \\
HH287 & 05:35:41.55 & -05:05:15.69 & MMS8            & 05:35:38.36 & -05:05:42.46 & [2] &  -60.8 & -13.3  & 47.5 \\
HH293 & 05:35:21.43 & -05:01:15.09 & MMS5            & 05:35:26.43 & -05:01:14.45 & [2] &   90.5 & -39.8  & 49.7 \\
HH294 & 05:35:22.97 & -05:04:02.20 & IRAS 05329-0505 & 05:35:26.38 & -05:03:53.45 & [3] &   99.7 &  12.8  & 86.9 \\
HH295 & 05:35:16.07 & -05:04:19.70 & IRAS 05329-0505 & 05:35:26.38 & -05:03:53.45 & [3] &   99.6 &  12.8  & 86.8 \\
HH331 & 05:35:08.24 & -05:00:48.13 & MMS2            & 05:35:18.24 & -05:00:34.86 & [2] &   95.1 &  46.6  & 48.5 \\
HH357 & 05:35:16.04 & -05:06:09.70 & MMS9            & 05:35:25.95 & -05:05:42.42 & [2] &  100.4 & -53.0  & 26.6 \\
HH383 & 05:35:26.40 & -05:07:52.46 & IRS1            & 05:35:24.55 & -05:10:28.62 & [4] &   -5.6 & -16.4  & 10.8 \\
HH384 & 05:35:25.37 & -05:09:23.38 & IRS2            & 05:35:26.37 & -05:09:25.75 & [4] &   81.0 & -17.1  & 81.9 \\
HH385 & 05:35:34.69 & -05:07:20.19 & MMS9            & 05:35:26.64 & -05:05:42.27 & [2] & -141.0 & -52.9  & 88.1 \\
&&&&&&&&&\\
\multicolumn{6}{l}{$\rho$ {\bf Ophiuchus}$\;\;$(SCUBA 850$\mu$m [5])} \\
HH312 & 16:25:55.95 & -24:20:46.97 & SR4             & 16:25:56.48 & -24:20:02.42 & [6] &   90.0 & -32.7 & 57.3 \\
HH313 & 16:26:18.82 & -24:23:06.44 & VLA1623         & 16:26:26.36 & -24:24:30.94 & [6] &   27.0 &  64.9 & 37.9 \\
&&&&&&&&&\\
\multicolumn{6}{l}{{\bf Serpens}$\;\;$(SCUBA 850$\mu$m [5])} \\
HH455 & 18:30:22.73 & 01:16:18.97 &   PS2            & 18:29:59.42 & 01:11:59.28  & [7] &  -55.5 & -7.4  & 48.1 \\
HH458 & 18:29:57.98 & 01:13:43.18 &  SMM3            & 18:29:59.18 &  01:13:58.26 & [7] &  132.5 & 15.2  & 62.7 \\
HH459 & 18:30:02.56 & 01:14:44.51 &  SMM3            & 18:29:59.18 & 01:13:58.26  & [7] &  -49.4 & 15.2  & 64.6 \\
HH460 & 18:29:38.59 & 01:18:23.78 & SMM1             & 18:29:49.75 &  01:15:18.58 & [7] &   42.3 & -41.7 & 84.0 \\
&&&&&&&&&\\

\multicolumn{6}{l}{$\rho$ {\bf L1524 (Taurus)}$\;\;$(SPITZER 160$\mu$m [8])} \\
HH156 & 04:18:51.54 & 28:20:28.14 &  CoKu Tau 1       & 04:18:51.60 & 28:20:28.14 & [9]  & 180.0  & 25.9  & 25.9 \\
HH157 & 04:22:00.89 & 26:57:37.61 &  Haro 6-5B        & 04:22:02.09 & 26:57:32.53 & [10] & -76.1  & 76.7  & 27.2 \\
HH390 & 04:19:40.85 & 27:15:52.86 & IRAS04166+2706    & 04:19:43.00 & 27:13:34.00 & [3]  &  02.6  & 43.7  & 41.1 \\
HH391 & 04:19:56.32 & 27:09:25.84 & IRAS04169+2702    & 04:19:59.20 & 27:09:59.00 & [3]  &  95.0  & 39.4  & 55.6 \\
&&&&&&&&&\\
\multicolumn{6}{l}{{\bf B18w(Taurus)}$\;\;$(SPITZER 160$\mu$m [8])} \\
HH184 & 04:29:23.66 & 24:33:01.44 & Haro 6-10        &  04:29:23.86 & 24:32:57.03 & [10] &  02.6  & 45.8  & 43.2 \\
HH300 & 04:25:23.04 & 24:23:20.11 & IRAS04239+2436   &  04:26:57.10 & 24:43:36.00 & [3]  & 139.4  & -42.5 & 01.9 \\
HH410 & 04:28:12.99 & 24:19:01.77 & Haro 6-10        &  04:29:23.86 & 24:32:57.03 & [10] & 130.3  & 45.8  & 84.5 \\
HH411 & 04:30:16.91 & 24:42:42.48 & Haro 6-10        &  04:29:23.86 & 24:32:57.03 & [3]  & -51.8  & 45.8  & 82.4 \\
HH412 & 04:29:52.97 & 24:37:10.08 & Haro 6-10        &  04:29:23.86 & 24:32:57.03 & [10] & -52.5  & 45.8  & 81.7 \\
HH413 & 04:29:52.99 & 24:38:12.08 & IRAS04264+2433   &  04:29:29.90 & 24:39:55.00 & [3]  & -51.5  & 45.8  & 82.7 \\
HH414 & 04:29:30.31 & 24:39:53.60 & IRAS04264+2433   &  04:29:29.90 & 24:39:55.00 & [3]  &   0.0  & 45.8  & 45.8 \\
HH466 & 04:33:35.42 & 24:21:32.09 & IRAS04305+2414   &  04:33:33.80 & 24:21:09.00 & [3]  & -64.0  & -66.1 & 02.1 \\
&&&&&&&&&\\
\multicolumn{6}{l}{{\bf L1521(Taurus)}$\;\;$(SPITZER 160$\mu$m [8])}\\
HH158 & 04:27:04.65 & 26:06:16.40 & DG Tau B          & 04:27:04.70 & 26:06:16.20 & [11] &   61.2 & 49.2  & 12.0 \\
HH159 & 04:27:02.05 & 26:05:41.58 & DG Tau B          & 04:27:04.70 & 26:06:16.20 & [11] &   85.6 & 49.2  & 36.4 \\
HH31  & 04:28:18.39 & 26:17:40.49 & IRAS04248+2612    & 04:27:57.12 & 26:19:17.00 & [3]  & -125.4 & -85.8 & 39.6 \\
&&&&&&&&&\\
\multicolumn{6}{l}{{\bf TMC(Taurus)}$\;\;$(SPITZER 160$\mu$m [8])}\\
HH395 & 04:40:08.73 & 25:46:44.44 & IRAS04369+2539    & 04:39:58.48 & 25:45:06.13 & [3]  &  -50.2 &   46.9 & 82.9 \\
HH408 & 04:41:38.92 & 25:56:26.27 & IRAS04385+2550    & 04:41:04.23 & 25:57:56.85 & [3]  & -103.8 & -143.8 & 40.0  \\
&&&&&&&&&\\
\multicolumn{6}{l}{{\bf NGC1333}$\;\;$(SCUBA 450$\mu$m [5])}\\
HH7   & 03:29:07.99 & 31:15:28.47 & SVS13             & 03:29:03.59 & 31:15:51.72 &  [12] & -113.7 & -41.1 & 72.6 \\
HH8   & 03:29:06.40 & 31:15:53.56 & SVS13             & 03:29:03.59 & 31:15:51.72 &  [12] & -116.6 & -41.1 & 75.5 \\
HH9   & 03:29:06.19 & 31:15:37.57 & SVS13             & 03:29:03.59 & 31:15:51.72 &  [12] &  -89.9 & -41.1 & 48.8 \\
HH10  & 03:29:05.29 & 31:15:46.63 & SVS13             & 03:29:03.59 & 31:15:51.72 &  [12] & -106.3 & -41.1 & 65.2 \\
HH11  & 03:29:04.49 & 31:15:53.67 & SVS13             & 03:29:03.59 & 31:15:51.72 &  [12] &  -89.9 & -41.1 & 48.8 \\
HH12  & 03:28:57.60 & 31:20:10.07 & SVS12             & 03:29:01.67 & 31:20:06.84 &  [12] &   90.0 & -20.3 & 69.7 \\
HH15  & 03:28:58.86 & 31:08:01.99 & IRAS03255+3103    & 03:28:37.11 & 31:13:28.30 &  [12] &  180.0 & -58.8 & 58.8 \\
HH338 & 03:28:12.82 & 31:19:43.63 & IRAS03255+3103    & 03:28:37.11 & 31:13:28.30 &  [13] &   56.3 & -58.8 & 64.9 \\
HH339 & 03:28:31.05 & 31:19:43.63 & IRAS03255+3103    & 03:28:37.11 & 31:13:28.30 &  [13] &   54.4 & -58.8 & 66.8 \\
HH340 & 03:28:44.94 & 31:05:38.79 & IRAS03256+3055    & 03:28:45.31 & 31:05:41.90 &  [13] &  180.0 & -58.8 & 58.8 \\
HH341 & 03:28:49.60 & 31:01:15.52 & NGC1333-IRAS2A    & 03:28:55.59 & 31:14:37.50 &  [13] &  180.0 & -58.8 & 58.8 \\
HH342 & 03:28:51.91 & 31:10:48.39 & IRAS03256+3055    & 03:28:45.31 & 31:05:41.90 &  [13] &  155.8 & -58.8 & 34.6 \\
HH343 & 03:28:54.40 & 31:05:21.25 & IRAS03256+3055    & 03:28:45.31 & 31:05:41.90 &  [13] &  180.0 & -58.8 & 58.8 \\
&&&&&&&&&\\
\hline
\end{tabular}
\end{center}
References. ---
[1] Nutter \& Ward-Thompson (2007);
[2] Chini \emph{et al} (1997);
[3] http://vizier.u-strasbg.fr/viz-bin/VizieR?-source=II/125;
[4] Yu \emph{et al} (1997);
[5] Di Francesco et al. (2008);
[6] Allen \emph{et al} (2002);
[7] Davis \emph{et al} (1999);
[8] Nutter \emph{et al} (2008); 
[9] Strom \emph{et al} (1986);
[10]Strom \& Strom (1994);
[11]Mundt \& Fried (1983);
[12]Strom \emph{et al} (1976);
[13]Davis \emph{et al} (2008).
\end{table*}

The systems we analyse are all in nearby star formation regions.

First, outflows are identified using the catalogue of Herbig Haro Objects (HHOs) by Reipurth (1999).

Second, for some of these HHOs, the driving YSO has been identified: in the Orion Integral Filament, by Chini et al. (1997) and Yu et al. (1997); in $\rho$ Ophiuchus, by Allen et al. (2002); in Serpens, by Davis et al. (1999); in Taurus, by Mundt \& Fried (1983), Strom et al. (1986) and Strom \& Strom (1994); and in NGC1333, by Strom, Vrba \& Strom (1976) and Davis et al. (2008). Knowing the positions of the HHO and the driving YSO, the direction of the outflow can be estimated.

Third we have used dust continuum maps to determine whether the driving YSO is embedded in a larger filamentary structure. For this purpose we have used SCUBA $850\,\mu{\rm m}$ data from Nutter \& Ward-Thompson (2007) for the Orion Integral Filament, SCUBA $850\,\mu{\rm m}$ data from the SCUBA Archive (Di Francesco et al. 2008) for $\rho$ Ophiuchus and Serpens; SCUBA $450\,\mu{\rm m}$ data from the SCUBA Archive (Di Francesco et al. 2008) for NGC 1333; and SPITZER $160\,\mu{\rm m}$ data from Nutter et al. (2008) for Taurus.

In the cases where filamentary structure can be identified, we have determined the direction of the filament, in as objective a manner as possible, using the SEXTRACTOR clump-finding algorithm in STARLINK. This algorithm requires the user to specify a detection threshold, in units of the background noise (for example ${\cal T}=6\sigma$) and an isophotal radius (e.g. ${\cal R}=2$), which determines the degree of smoothing over neighbouring pixels. Maxima in the map are fitted with elliptical isophotes, and the semi-major axis of the ellipse then corresponds to the direction of the filament. We vary ${\cal T}$ and ${\cal R}$ until this direction is approximately idependent of the exact choice of ${\cal T}$ and ${\cal R}$, {\it and} the direction corresponds reasonably with what the human eye sees.

If $\eta_{_{\rm OUT}}$ is the direction of the outflow, and $\eta_{_{\rm FIL}}$ is the direction of the filament (both measured clockwise from north), then the projected angle between outflow and filament is given by
\begin{eqnarray}
\gamma&=&{\rm MIN}\,\left\{\,|\eta_{_{\rm OUT}}-\eta_{_{\rm FIL}}|\,,\,\pi-|\eta_{_{\rm OUT}}-\eta_{_{\rm FIL}}|\,\right\}
\end{eqnarray}

In Table \ref{TAB:DATA} we list, in columns 1 to 3, the name, right ascension and declination for the HHOs used in our analyis; these are all from Reipurth (1999). In columns 4 to 7, we give the name, right ascension and declination for the driving YSO, plus the reference for the identification of the driving YSO. In columns 8 to 10, we give the angles $\eta_{_{\rm OUT}}$, $\eta_{_{\rm FIL}}$ and $\gamma$. The systems (filament, YSO and HH object/outflow) are grouped in eight different star formation regions, and the source and wavelength of the continuum map used to define the filaments is given after the name of the star formation region.

\begin{figure}
\centering
\includegraphics[angle=0,width=9.0cm]{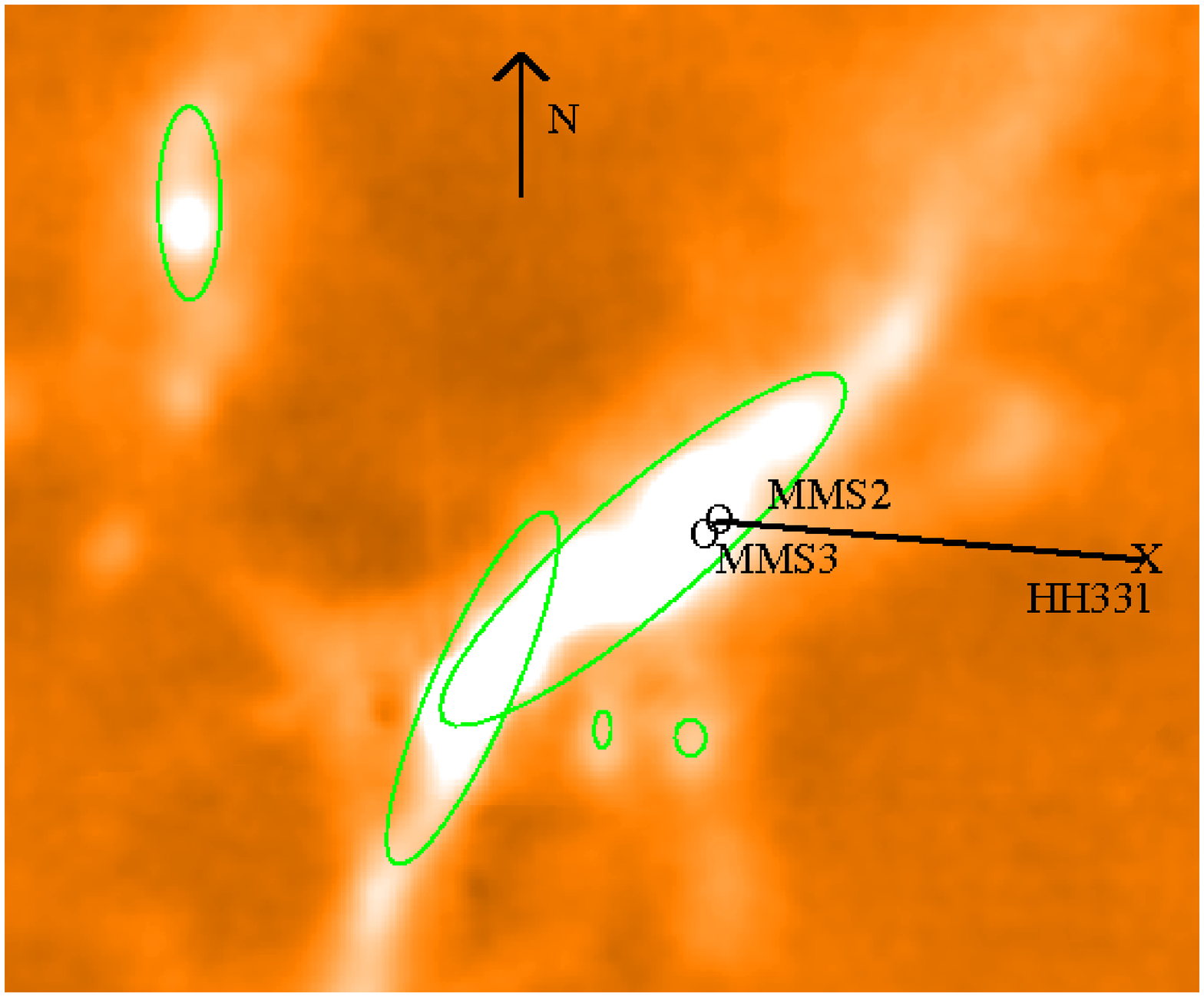}
\caption{False-colour map of the 850 $\mu$m continuum emission from part of the Orion Integral Filament (Nutter \& Ward-Thompson, 2007), showing the positions of MMS2, MMS3, HH331, and the ellipse fitted to the local portion of the filament in order to estimate its direction.}
\label{FIG:OIF}
\end{figure}

Fig. \ref{FIG:OIF} shows the continuum maps used to identify filaments; the long lines with arrows at either end indicate the derived directions of the underlying filaments. The circles indicate the positions of YSOs, and the crosses indicate the positions of HH Objects; the lines joining these are the inferred directions of the outflows. 

\section{The distribution of projected angles between filaments and outflows}

Without loss of generality, we define Cartesian axes $(x,y,z)$ with unit vectors $(\hat{\bf i},\hat{\bf j},\hat{\bf k})$, oriented so that (i) the filament is along the $z$-axis, i.e. it has a unit vector
\begin{eqnarray}
\hat{\bf n}_{_{\rm FIL}}&=&\hat{\bf k}\,;
\end{eqnarray}
and (ii) the outflow is in the $(y,z)$-plane. If the outflow makes an angle $\psi$ with the $y$-axis, then the outflow has a unit vector
\begin{eqnarray}
\hat{\bf n}_{_{\rm OUT}}&=&\cos(\psi)\hat{\bf j}\,+\,\sin(\psi)\hat{\bf k}\,.
\end{eqnarray}

In order to compute the projected (i.e. observed) angle, $\gamma$, between the filament and the outflow, we assume that the observer is located at large distance (compared with the size of the filament/outflow system) in the direction of the unit vector
\begin{eqnarray}
\hat{\bf n}_{_{\rm OBS}}&=&\sin(\theta)\cos(\phi)\hat{\bf i}\,+\,\sin(\theta)\sin(\phi)\hat{\bf j}\,+\,\cos(\theta)\hat{\bf k}\,.
\end{eqnarray}

We can construct a two-dimensional Cartesian co-ordinate system on the observer's sky using the unit vectors
\begin{eqnarray}
\hat{\bf a}&=&\frac{\hat{\bf k}\vm\hat{\bf n}_{_{\rm OBS}}}{|\hat{\bf k}\vm\hat{\bf n}_{_{\rm OBS}}|}\;=\;-\,\sin(\phi)\hat{\bf i}\,+\,\cos(\phi)\hat{\bf j}\,,\\\nonumber
\hat{\bf b}&=&\hat{\bf n}_{_{\rm OBS}}{\vm}\,\hat{\bf a}\;=\;-\cos(\theta)\cos(\phi)\hat{\bf i}\,-\,\cos(\theta)\sin(\phi)\hat{\bf j}\,+\,\sin(\theta)\hat{\bf k}\,;
\end{eqnarray}
the orientation of these axes is such that $\hat{\bf a}\vm\hat{\bf b}=\hat{\bf n}_{_{\rm OBS}}$, so that (notionally) $\hat{\bf a}$ is the abscissa, and $\hat{\bf b}$ is the ordinate.

On the observer's sky, the unit vector along the filament is then
\begin{eqnarray}\nonumber
\hat{\bf n}'_{_{\rm FIL}}&=&\frac{(\hat{\bf n}_{_{\rm FIL}}\cdot\hat{\bf a})\hat{\bf a}+(\hat{\bf n}_{_{\rm FIL}}\cdot\hat{\bf b})\hat{\bf b}}{|(\hat{\bf n}_{_{\rm FIL}}\cdot\hat{\bf a})\hat{\bf a}+(\hat{\bf n}_{_{\rm FIL}}\cdot\hat{\bf b})\hat{\bf b}|}\;\,=\;\,\hat{\bf b}\,,
\end{eqnarray}
and the unit vector along the outflow is
\begin{eqnarray}\nonumber
\hat{\bf n}'_{_{\rm OUT}}\!&\!=\!&\frac{(\hat{\bf n}_{_{\rm OUT}}\cdot\hat{\bf a})\hat{\bf a}+(\hat{\bf n}_{_{\rm OUT}}\cdot\hat{\bf b})\hat{\bf b}}{|(\hat{\bf n}_{_{\rm OUT}}\cdot\hat{\bf a})\hat{\bf a}+(\hat{\bf n}_{_{\rm OUT}}\cdot\hat{\bf b})\hat{\bf b}|}\\\nonumber
\!&\!=\!&\frac{\cos(\psi)\cos(\phi)\hat{\bf a}+\left\{\sin(\psi)\sin(\theta)\!-\!\cos(\psi)\cos(\theta)\sin(\phi)\right\}\hat{\bf b}}{f^{1/2}(\psi,\theta,\phi)}\,,\\
\end{eqnarray} 
where
\begin{eqnarray}\nonumber
f(\psi,\theta,\phi)\!&\!=\!&\!\cos^2(\psi)\cos^2(\phi)+\cos^2(\psi)\cos^2(\theta)\sin^2(\phi)\\\nonumber
&&\!+\!\sin^2(\psi)\sin^2(\theta)\!-\!2\cos(\psi)\sin(\psi)\cos(\theta)\sin(\theta)\sin(\phi)\\
\end{eqnarray}

Hence the projected angle, $\gamma$, between the filament and the outflow, is given by
\begin{eqnarray}\nonumber
\cos(\gamma)&=&\hat{\bf n}'_{_{\rm FIL}}\cdot\hat{\bf n}'_{_{\rm OUT}}\\
&=&\frac{-\cos(\psi)\cos(\theta)\sin(\phi)+\sin(\psi)\sin(\theta)}{f^{1/2}(\psi,\theta,\phi)}\,.
\end{eqnarray}

We shall assume that $\psi$ is distributed according to the truncated Gaussian
\begin{eqnarray}
p_{_{\psi}}\,d\psi&=&\frac{C}{(2\pi)^{1/2}\sigma_{\!\psi}}\,\exp\left\{-\,\frac{\psi^2}{2\sigma_{\!\psi}^2}\right\}\,d\psi\,,\hspace{1.0cm}|\psi|\leq \frac{\pi}{2}\,,
\end{eqnarray}
where
\begin{eqnarray}
C^{-1}&=&\int_{_{\psi=-\pi/2}}^{^{\psi=+\pi/2}}\,\frac{1}{(2\pi)^{1/2}\sigma_{\!\psi}}\,\exp\left\{-\,\frac{\psi^2}{2\sigma_{\!\psi}^2}\right\}\,d\psi\,.
\end{eqnarray}

If there are no selection effects, and none of the observed outflows are correlated in direction, then -- again without loss of generality - putative observers can be distributed isotropically over one octant of the sky, with
\begin{eqnarray}
p_{\theta\,\phi}\,d\theta\,d\phi&=&\sin(\theta)\,d\theta\;\frac{2\,d\phi}{\pi}\,,\hspace{1.0cm}0\leq\theta,\phi\leq\frac{\pi}{2}\,.
\end{eqnarray}

For any given $\sigma_{\!\psi}$ (measuring the dispersion of the outflow direction about the orthogonal to the filament), the cumulative distribution of projected angles can be computed by means of a Monte Carlo integration. To do this we consider many independent sequences, each of ${\cal N}=45$ randomly distributed sample points in the integration space $(\psi,\theta,\phi)$. In this way we obtain both the expected cumulative distribution, $P(\gamma)$, and a measure of the dispersion about $P(\gamma)$ for a finite sample of ${\cal N}$ observations.

Specifically, we record the $n^{\rm th}\;$ $\gamma$-value from each sequence, and then compute their mean, $\mu_{\,\gamma}(n)$, and their standard deviation, $\sigma_{\gamma}(n)$. The expected cumulative distribution, $P(\gamma)$, is then a plot of $\left(n-\frac{1}{2}\right)/{\cal N}$ against $\mu_{\gamma}(n)$. The $1\sigma$ dispersion about this distribution is represented by the plot of $\left(n-\frac{1}{2}\right)/{\cal N}$ against $\mu_{\,\gamma}(n)\pm\sigma_{\gamma}(n)$. The distributions obtained in this way for various different $\sigma_{\!\psi}$ are presented, and compared with the data, in Fig. 1.



\end{document}